\documentclass[conference]{IEEEtran}

\usepackage{amssymb}
\usepackage{color}
\usepackage{cite}
\usepackage{graphicx,epsfig}
\usepackage[cmex10]{amsmath}
\usepackage{subfig}

\begin{document}

\title{Performance Analysis of Self-Interference~Cancellation in Full-Duplex~Large-Scale~MIMO~Systems}

\author{\IEEEauthorblockN{Yeon-Geun Lim\IEEEauthorrefmark{1},
Daesik Hong\IEEEauthorrefmark{2},
and Chan-Byoung Chae\IEEEauthorrefmark{1}}
\IEEEauthorblockA{\IEEEauthorrefmark{1}School of Integrated Technology,
\IEEEauthorrefmark{2}School of Electrical and Electronic Engineering\\
Yonsei University, Korea\\
Email: \{yglim, daesikh, cbchae\}@yonsei.ac.kr\\}}

\maketitle

\begin{abstract}
This paper presents the performance analysis of the self-interference cancellation (SIC) methods in full-duplex large-scale multiple-input multiple-output systems. To mitigate self-interference (SI), we assume that the full duplex-base station (BS) uses SI-subtraction or spatial suppression. Analytical and numerical results confirm that the SI-subtraction outperforms the spatial suppression for SIC in a perfect channel estimation case. It is also concluded that the uplink and overall ergodic rates performance of the spatial suppression is respectively better than those of the SI-subtraction in a imperfect channel estimation case under a given system constraint such as uplink/downlink sum rates and the total transmit power at the BS.\\
\end{abstract}


\IEEEpeerreviewmaketitle

\section{Introduction}

By the year 2020, it is expected that mobile data traffic will have grown a thousand-fold (1000x). To support such tremendous data traffic, researchers should consider in a next generation communication system such key issues to increasing network capacity (bps/area) as high spectral efficiency (bps/Hz), spectrum extension (Hz/cell), and high network density (cell/area) should be jointly considered. Over the last few decades, multiple-input multiple-output (MIMO) wireless communication techniques have evolved for high spectral efficiency~\cite{Chae_SPMag_07}. Full-duplex transmission has also been studied for double spectrum extension \cite{Sachin_fd,FD_Hong15}. For high network density, researchers have considered deploying a large number of small cells.


Recently, to maximize spectral efficiency and to conserve energy, the researchers in \cite{Marzetta_WC10,Yang, Rusek_SPMag_12, Ngo_TCOM12, Lim_TWC} have proposed large-scale MIMO systems. They considered simple linear precoders and filters to mitigate the interference that arise from large-scale MIMO transmission.
The authors in \cite{Yang, Rusek_SPMag_12} studied the downlink performance of maximum ratio transmission (MRT) and zero-forcing (ZF) precoders while assuming perfect and imperfect channel estimations at the base station (BS). 
The authors in \cite{Ngo_TCOM12} investigated the uplink performance of maximum ratio combining (MRC), ZF and minimum mean square error (MMSE) filters while assuming perfect and imperfect channel estimations at the BS. They showed that the transmit energy of the users could be conserved by the power-scaling law when the BS with a large-scale array serves a small number of users that equipped with one antenna. 
The authors in \cite{Lim_TWC} analyzed both downlink and uplink performances of the cell-boundary users (including the performance analysis of a large number of users) considering the impact of the low transmit or received power. They also proposed downlink precoding normalization methods and transceiver mode selection algorithms when the BS uses the MRT/ZF precoder at downlink and the MRC/ZF filter at uplink.


Another relevant technology is the full duplex transmission that can double the spectrum bandwidth compared to half-duplex transmission, such as time division duplex (TDD) and frequency division duplex (FDD). In contrast to the half duplex transmission, the self-interference (SI) which is a transmitted downlink signal and directly received at the BS while receiving a uplink signal should be mitigated. The authors in \cite{FDSI_TSP11} compared the self-interference cancellation (SIC) by SI-subtraction and that by spatial suppression in a MIMO-relay system with a small number of antennas (up to 4 antennas) at the BS and the user. They showed the performance of the spatial suppression was better than that of the SI-subtraction but depended on the rank of the SI channel and that it required additional antennas. In a practical system with a small number of antennas at the BS, the SI-subtraction is a widely considered the SIC technique due to the limitation of the requirement of additional antennas. The researchers in \cite{MK_ComMag15} prototyped SI-subtraction-based algorithm on a software-defined radio (SDR) platform in real time. They combined a dual-polarization antenna-based analog cancellation with the SI-subtraction-based algorithm and achieved 1.9 times higher spectral efficiency than the half duplex transmission.


There are some integrated techniques for high spectral efficiency, spectrum extension, and high network density such as a large-scale MIMO system with a small cell technique and full duplex transmission with a large-scale antenna array, which can support huge data traffic demands. The researchers in \cite{Jang_WCM} presented a real time three-dimensional hybrid beamforming for next generation communication systems, which is equipped with the large-scale antenna array at a small cell. \cite{Lim_ICC14,Sim_JCN} proposed a compressed channel feedback for a highly correlated channel that is yielded from compact antenna arrays at the small BS. Similarly,  \cite{Kwon_TMTT} proposed a novel channel feedback for a radio frequency lens-embedded large-scale MIMO system, and analyzed the performance at a small cell. The authors in \cite{SIM_MCOM} analyzed the performance of the nonlinear SIC techniques for a full duplex system at a small cell. \cite{Yin_asi13} proposed the spatial suppression-based algorithms such as extended ZF and extended MMSE precoders for full duplex large-scale MIMO systems. Since their approximations of the downlink/uplink sum rate are based on an interference-free channel, their analysis lack certitude. Moreover, the authors ignored the impact of imperfect channel estimation--the key issue in SIC problem. 


In this paper, we present a performance analysis of the SI-subtraction and the spatial suppression in full duplex large-scale MIMO systems. The spatial suppression could be a good option of the SIC in large-scale MIMO systems because many additional antennas can be utilized. From the analysis and the numerical results, we investigate the SIC algorithm which is better.


\section{System Model}

\begin{figure}[t]
 \centerline{\resizebox{0.9\columnwidth}{!}{\includegraphics{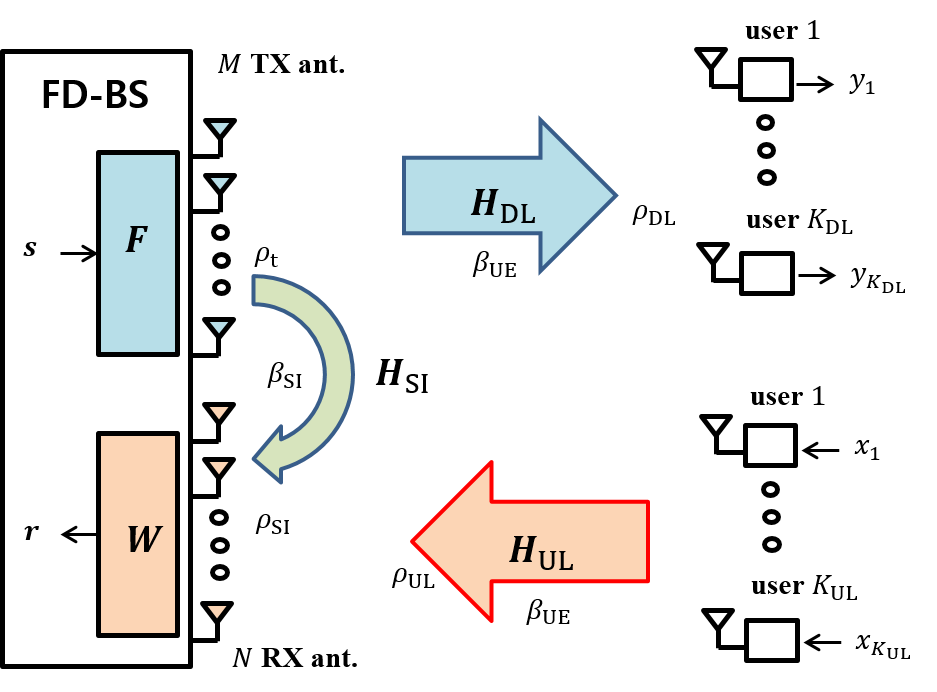}}}
  \caption{A block diagram of a full duplex multi-user large-scale MIMO system.}
  \label{Sysmodel}
\end{figure}

In this section, we introduce the basic notation used in this paper and the full duplex large-scale MIMO system.\footnote{Throughout this paper, we use upper and lower case boldfaces to describe matrix $\pmb{A}$ and vector $\pmb{a}$, respectively. We denote the inverse, transpose and the Hermitian of matrix $\pmb{A}$ by $\pmb{A}^{-1}$, $\pmb{A}^{T}$, and $\pmb{A}^{*}$, respectively.}  

\subsection{System Model: Full Duplex BS with Large-Scale Arrays}

Consider a full duplex multi-user MIMO channel, as illustrated in Fig. \ref{Sysmodel}, with $M(\ge N+K)$ transmit antennas and $N$ receive antennas at the BS (a total number of antennas at the BS is $L$, i.e., $L=M+N$) that serves $K$ downlink users and $K$ uplink users equipped with one antenna for each. This asymmetric antenna system ensures the small and low cost BS by reducing radio frequency chains that include power amplifiers ($N\ge M+K$) at a transmitter or low-noise amplifiers ($M\ge N+K$) at a receiver in the fixed size of total number of radio frequency chains (i.e., $L$). We assume the BS has more transmit antennas than receive antennas to support the asymmetric data traffic load which the downlink has more offered than the uplink in practice. 
We also assume that there is no user-to-user interference where the path-losses between the users are sufficiently large. The free-space propagation gain $\beta_\text{SI}$ is considered at the SI channel.
The total transmit signal-to-noise ratio (SNR) at the BS, the received downlink SNR at the users, the received uplink SNR at the BS, and the received SI SNR at the BS are respectively denoted by $\rho_\text{t}(=\text{P}_\text{t}/\text{P}_\text{n})$, $\rho_{_\text{DL}}(=\rho_{t}\beta_\text{UE})$, $\rho_{_\text{UL}}$, and $\rho_{_\text{SI}}(=\rho_{t}\beta_\text{SI})$, where $\text{P}_\text{t}$, $\text{P}_\text{n}$, and $\beta_\text{UE}$ are the total transmit power, the noise power of the additive white Gaussian noise of the channel, and the propagation gain between the BS and the user respectively.\footnote{There is a trade-off between a downlink rate and a uplink rate according to $\rho_\text{t}$ if the values of the propagation gains are fixed. To simplify this effect in the expressions of the signal-to-interference-noise-ratio (SINR), we define the total transmit, received downlink, uplink, and SI SNRs. We also assume the distances from the BS to all users are the same and there are no shadow fading for a simplicity.}
We assume that the SI, downlink and uplink channels are Rayleigh channel modeled by the $N\times M$ matrix $\pmb{H}_\text{SI}$, the $K\times M$ matrix $\pmb{H}_\text{DL}$, and the $N\times K$ matrix $\pmb{H}_\text{UL}$, respectively, the elements of which are i.i.d. (independent identically distributed) complex Gaussian random variables with zero mean and unit variance. 

Let $\pmb{g}_k$ denote the $k$-th column vector of transmit precoding matrix $\pmb{G}$ and $s_k$ represent the transmit symbol for the $k$-th user at downlink. 
Similarly, let $\pmb{w}_k$ denote the $k$-th column vector of receive combining matrix $\pmb{W}$ and $x_k$ represent the transmit symbol for the $k$-th user at uplink.
Also, let $\pmb{n}_k$ be the additive white Gaussian noise vector of the $k$-th user. 
The received signal at the $k$-th user is then expressed by
\begin{align}
y_k=\underbrace{\sqrt{\rho_{_\text{DL}}}\pmb{h}_{\text{DL}, k}^T\pmb{g}_ks_k}_{\text{desired signal}}+\underbrace{\sqrt{\rho_{_\text{DL}}}\sum_{\ell=1,\ell\neq k}^{K}\pmb{h}_{\text{DL}, k}^T\pmb{g}_{\ell}s_\ell}_{\text{user-interference}}+n_{_\text{DL}, k}\label{eq_rcv_dl}
\end{align}
where $\pmb{h}_{\text{DL}, k}$ denotes the the $k$-th column vector of $\pmb{H}_\text{DL}^T$.
Also, the received signal for the $k$-th user at the BS is expressed by
\begin{align}
r_k=\underbrace{\sqrt{\rho_{_\text{UL}}}\pmb{w}_k^T\pmb{h}_{\text{UL}, k}x_k}_{\text{desired signal}}+\underbrace{\sqrt{\rho_{_\text{UL}}}\sum_{\ell=1,\ell\neq k}^{K}\pmb{w}_k^T\pmb{h}_{\text{UL}, \ell}x_\ell}_{\text{user-interference}}\nonumber\\
+\underbrace{\sqrt{\rho_{_\text{SI}}/\alpha_\text{anc}}\pmb{w}_k^T\pmb{H}_\text{SI}\pmb{G}\pmb{s}}_{\text{self-interference}}+\pmb{w}_k^Tn_{_\text{UL}, k}\label{eq_rcv_ul}
\end{align}
where $\pmb{h}_{\text{UL}, k}$ and $\alpha_\text{anc}$ denotes the $k$-th column vector of $\pmb{H}_\text{UL}$ and the level of the passive analog SIC. To satisfy a power constraint, the normalized transmit beamforming vectors (the columns of the precoding matrix) with vector normalization are given as $\pmb{g}_k=\pmb{f}_k/(\sqrt{K}||\pmb{f}_k||)$ \cite{Lim_TWC}.

From (\ref{eq_rcv_dl}) and (\ref{eq_rcv_ul}), we can derive the SINR of the $k$-th user at downlink and uplink, respectively, as given by
\begin{align}
{\gamma}_{k}^\text{DL}&=\frac{\rho_{_\text{DL}}\left|\pmb{h}_{\text{DL}, k}^T\frac{\pmb{f}_k}{\sqrt{K}||\pmb{f}_k||}\right|^2}{\rho_{_\text{DL}}\sum_{\ell=1,\ell\neq k}^{K}\left|\pmb{h}_{\text{DL}, k}^T\frac{\pmb{f}_{\ell}}{\sqrt{K}||\pmb{f}_{\ell}||}\right|^2+1},\label{SINR_DL}\\
{\gamma}_{k}^\text{UL}&=\frac{\rho_{_\text{UL}}|\pmb{w}_k^T\pmb{h}_{\text{UL}, k}|^2}{\rho_{_\text{UL}}\sum_{\ell=1,\ell\neq k}^{K}|\pmb{w}_k^T\pmb{h}_{\text{UL}, \ell}|^2+\frac{\rho_{_\text{SI}}}{\alpha_\text{anc}}\Omega+||\pmb{w}_k||^2}\label{SINR_UL}
\end{align} where $\Omega=||\pmb{w}_k^T\pmb{H}_\text{SI}\pmb{G}||^2$.

\subsection{Precoding/Receive Combining Matrix Design and Self-Interference Cancellation Methods} 
Let $\pmb{\hat H}_{\text{DL}}$, $\pmb{\hat H}_{\text{UL}}$ and $\pmb{\hat H}_{\text{SI}}$ be the estimated channel matrices of $\pmb{H}_{\text{DL}}$, $\pmb{H}_{\text{UL}}$ and $\pmb{H}_{\text{SI}}$. Since the full duplex-BS always perfectly knows its precoding/receive matrix and transmitted downlink symbols, it can mitigate SI by directly subtracting $\sqrt{\rho_{_\text{SI}}}\pmb{w}_k^T\hat{\pmb{H}}_\text{SI}\pmb{G}\pmb{s}$, of which the SIC is referred to as the SI-subtraction. The received signal after SI-subtraction is
\begin{align}
r_{\text{stt}, k}=r_k-\sqrt{\rho_{_\text{SI}}/\alpha_\text{anc}}\pmb{w}_k^T\hat{\pmb{H}}_\text{SI}\pmb{G}\pmb{s}.\nonumber
\end{align}
A ZF precoder for downlink is assumed at the BS, in the cases of the the full duplex-BS with the SI-subtraction and without the SIC:
\begin{align}
\pmb{F}_\text{ZF}=\pmb{\hat H}_\text{DL}^*(\pmb{\hat H}_\text{DL}\pmb{\hat H}_\text{DL}^*)^{-1}.\nonumber
\end{align}
From the antenna configuration of $M\ge N+K$, the additional antennas for the spatial suppression can be utilized. The spatial suppression-based algorithms for a large-scale MIMO system was proposed in \cite{Yin_asi13}. 
The spatial suppression-based precoder is given~by
\begin{align}
\pmb{F}_\text{sps}=\pmb{\hat H}_\text{ext}^*(\pmb{\hat H}_\text{ext}\pmb{\hat H}_\text{ext}^*)^{-1},~~~~\pmb{\hat H}_\text{ext}=\begin{bmatrix}\pmb{\hat H}_\text{DL}\\ \pmb{\hat H}_\text{SI}\end{bmatrix}.\nonumber
\end{align}
To eliminate user-interference signals at uplink, we use following ZF receive combining matrix in both transceiver with the SI-subtraction/spatial suppression and without the SIC:
\begin{align}
\pmb{W}=({\pmb{\hat H}}_\text{UL}^*{\pmb{\hat H}}_\text{UL})^{-1}{\pmb{\hat H}^*_\text{UL}}.\nonumber
\end{align}

\subsection{Estimated Channel Model}

We assume pilot symbols are orthogonally transmitted under TDD-based transmission. The number of pilot symbols is $K$ and $N$ at the downlink/uplink channel and the SI channel. Since estimation errors are independent with original channels, estimated channels can be modeled by estimated error matrices ($\pmb{E}$) that have i.i.d. $\mathcal{CN}(0,\epsilon^2)$ elements as follows:
\begin{align}
\pmb{\hat H}_\text{DL/UL/SI}=\pmb{H}_\text{DL/UL/SI}+\pmb{E}_{_\text{DL/UL/SI}}.\label{est_error}
\end{align}
We assume that MMSE-based channel estimation is used at the BS \cite{Ngo_TCOM12}.
\subsubsection{SI Channel Estimation Error}\label{channel_model2} If the hardware of the full duplex system is perfect, the MMSE-based estimation error of the SI channel could be close to zero due to high received pilot power. In practice, the estimation error of the SI channel yields residual SI that impact on the significant performance degradation of the full duplex systems which error is imperfect because of hardware impairments such as the nonlinearity of the power amplifier, a quantization error, and phase noise \cite{Day_TSP}. Thus, it is difficult to model the estimated SI channel while considering the hardware impairments. The variance of the estimation error is easily modeled as the fixed value of the normalized-MSE (NMSE) between the estimated channel and the original channel, which depends on the performance of the estimation algorithm and the specification of the hardware. The variance of the estimation error of the SI channel is given by \begin{align}\epsilon_{_\text{SI}}^2=\text{NMSE}\nonumber.\end{align}


\subsubsection{Downlink/uplink Channel Estimation Error} In the case of the downlink/uplink channel estimation, we assume there are no error terms from the hardware impairments since the distance between the BS and the users are sufficiently large, i.e., the path-loss is large, which means the errors from the hardware impairments become relatively smaller than the additive white Gaussian noise. Thus, the MMSE estimation error can be considered \cite{Ngo_TCOM12}. The variance of the estimation error for the downlink/uplink channel is given by
\begin{align}
\epsilon_{_\text{DL/UL}}^2=\frac{\beta_\text{UE}}{K\rho_{_\text{u}}\beta_\text{UE}+1}\label{ULDL_error}
\end{align}
where $\rho_{_\text{u}}$ is the transmit SNR at the user.
Note that any estimation algorithm dependent on the transmit pilot power can be modeled similarly.

\section{Performance Analysis}
In this section we present the performance analysis of the SI-subtraction and the spatial suppression in full duplex large-scale MIMO systems. 
\subsection{Perfect Channel Estimation}
Let ${\bar \Omega}=\mathbb{E}\{\Omega\}$ be expected SI power. The SI-subtraction and the spatial suppression transceivers with perfect channel estimation promise ${\bar \Omega}=0$. If there is no SIC at the BS ($\alpha_\text{anc}=0$), ${\bar \Omega}=\mathbb{E}\left\{||\pmb{w}_k||^2\right\}$ because $\pmb{H}_\text{SI}$ and $\pmb{G}$ are independent \cite{Yin_asi13}. Thanks to the approximation methods of \cite{Lim_TWC}, we can derive the approximations of the ergodic sum rate of the full duplex transceiver without the SIC, with the SI-subtraction, and with the spatial suppression, respectively, as follows:  
\begin{align}
\begin{split}
\mathcal{R}_\text{woSIC}=K\log_2\left\{1+\frac{\rho_{_\text{DL}}(M-K+1)}{K}\right\}\\
+K\log_2\left\{1+\frac{\rho_{_\text{UL}}(N-K+1)}{\rho_{_\text{SI}}+1}\right\},\label{rate_noSIC}
\end{split}
\end{align}
\begin{align}
\begin{split}
\mathcal{R}_\text{stt}=K\log_2\left\{1+\frac{\rho_{_\text{DL}}(M-K+1)}{K}\right\}\\
+K\log_2\left\{1+\rho_{_\text{UL}}(N-K+1)\right\},
\end{split}\label{rate_stt}
\end{align}
\begin{align}
\begin{split}
\mathcal{R}_\text{sps}=K\log_2\left\{1+\frac{\rho_{_\text{DL}}(M-N-K+1)}{K}\right\}\\
+K\log_2\left\{1+\rho_{_\text{UL}}(N-K+1)\right\}
\end{split}\label{rate_sps}
\end{align}
where the approximations of the ergodic sum rate at downlink result from $\mathbb{E}\left\{\frac{1}{||\pmb{f}_{k,\text{ZF}}||^2}\right\}=M-K+1$ and $\mathbb{E}\left\{\frac{1}{||\pmb{f}_{k,\text{sps}}||^2}\right\}=M-(N+K)+1$ as well as those at uplink are derived from $\mathbb{E}\left\{\frac{1}{||\pmb{w}_k||^2}\right\}=N-K+1$.
The approximations of the ergodic sum rate of the half duplex with ZF precoding and ZF receive combining is the same as $\frac{1}{2}\mathcal{R}_\text{stt}$ \cite{Lim_TWC}.

\subsection{Imperfect Channel Estimation}

In this section, we analyze uplink SINR, which could be a more meaningful element than downlink SINR in the performance perspective at a full duplex system because of SI.  From (\ref{SINR_UL}) and (\ref{est_error}), the SINR of the $k$-th user at uplink in a full duplex system with imperfect channel estimation and the ZF receiver at the BS is given by
\begin{align}
{\gamma}_{\text{ZF}, k}^\text{UL}&=\frac{\rho_{_\text{UL}}|\pmb{w}_k^T\pmb{\hat{h}}_{\text{UL}, k}|^2}{\rho_{_\text{UL}}I+\rho_{_\text{SI}}\Omega+||\pmb{w}_k||^2}\nonumber\\
&=\frac{\rho_{_\text{UL}}}{\rho_{_\text{UL}}\sum_{\ell=1}^{K}|\pmb{w}_k^T\pmb{e}_{\text{UL},\ell}|^2+\frac{\rho_{_\text{SI}}}{\alpha_\text{anc}}\Omega+||\pmb{w}_k||^2}\label{SINR_UL_imperfect}
\end{align}
where $\pmb{e}_{\text{UL}, k}$ denotes the $k$-th column vector of $\pmb{E}_\text{UL}$, $I=\sum_{\ell=1,\ell\neq k}^{K}|\pmb{w}_k^T\pmb{\hat{h}}_{\text{UL}, \ell}|^2+\sum_{\ell=1}^{K}|\pmb{w}_k^T\pmb{e}_{\text{UL},\ell}|^2$, $|\pmb{w}_k^T\pmb{\hat{h}}_{\text{UL}, k}|^2=1$, and $|\pmb{w}_k^T\pmb{\hat{h}}_{\text{UL}, \ell}|^2=0$, respectively.

\subsubsection{SINR of a full duplex receiver without the SIC at uplink}
By using the approximation methdos of \cite{Ngo_TCOM12, Lim_TWC}, the approximation of the uplink SINR without the SIC is derived from (\ref{ULDL_error}), (\ref{SINR_UL_imperfect}) and ${\bar \Omega}=\mathbb{E}\left\{||\pmb{w}_k||^2\right\}$  as follows:
\begin{align}
{\gamma}_{\text{woSIC}, k}^\text{UL}=\frac{K\rho_{_\text{UL}}^2(N-K)}{2K\rho_{_\text{UL}}+\frac{\rho_{_\text{SI}}}{\alpha_\text{anc}}(K\rho_{_\text{UL}}+1)+1}.\label{SINR_noSIC}
\end{align}

\subsubsection{SINR of SI-Subtraction at Uplink}
The expected SI power of the full duplex transceiver with the SI-subtraction is
\begin{align}
{\bar \Omega}_\text{stt}
&=\mathbb{E}\left\{||\pmb{w}_k^T(\pmb{H}_\text{SI}-\pmb{\hat H}_\text{SI})\pmb{G}||^2\right\}\nonumber\\
&=\mathbb{E}\left\{||\pmb{w}_k^T\pmb{E}\pmb{G}||^2\right\}\nonumber\\
&\overset{(a)}{=}\epsilon_{_\text{SI}}^2\mathbb{E}\left\{||\pmb{w}_k||^2\right\}\label{stt_SI}
\end{align}
where ($a$) results from \cite{Yin_asi13} since $\pmb{E}$ and $\pmb{G}$ are independent. From (\ref{SINR_UL_imperfect}) and (\ref{stt_SI}), we can derive the approximation of the uplink SINR with the SI-subtraction, which is referred as
\begin{align}
{\gamma}_{\text{stt}, k}^\text{UL}=\frac{K\rho_{_\text{UL}}^2(N-K)}{2K\rho_{_\text{UL}}+\frac{\rho_{_\text{SI}}}{\alpha_\text{anc}}\epsilon_{_\text{SI}}^2(K\rho_{_\text{UL}}+1)+1}.\label{SINR_stt}
\end{align}

\subsubsection{SINR of Spatial Suppression at Uplink}
Since $\pmb{H}_\text{SI}$ and $\pmb{G}_\text{sps}$ are not independent in the case of the spatial suppression, we should consider the best and the worst estimation error to make the independent condition. In the case of the best channel estimation error ($\epsilon_{_\text{SI}}\rightarrow0$), the estimated SI channel tends to the original SI channel ($\pmb{\hat H}_{\text{SI}}\rightarrow\pmb{H}_{\text{SI}}$) so $\pmb{E}$ and $\pmb{G}_\text{sps}$ tend to be independent. Thus, the expected SI power of the full duplex transceiver with the spatial suppression (the best case) is given by
\begin{align}
{\bar \Omega}_{\text{sps}(\epsilon_{_\text{SI}}\rightarrow0)}
&=\mathbb{E}\left\{||\pmb{w}_k^T\pmb{H}_\text{SI}\pmb{G}_\text{sps}||^2\right\}\nonumber\\
&=\mathbb{E}\left\{||\pmb{w}_k^T(\pmb{\hat H}_\text{SI}-\pmb{E})\pmb{G}_\text{sps}||^2\right\}\nonumber\\
&=\mathbb{E}\left\{||\pmb{w}_k^T\pmb{E}\pmb{G}_\text{sps}||^2\right\}\nonumber\\
&=\epsilon_{_\text{SI}}^2\mathbb{E}\left\{||\pmb{w}_k||^2\right\}\label{sps_SI_b}
\end{align}
where $\pmb{\hat H}_\text{SI}\pmb{G}_\text{sps}=\pmb{0}$. Similarly, in the case of the worst channel estimation error ($\epsilon_{_\text{SI}}\rightarrow\infty$), the estimated SI channel tends to the channel estimation error matrix ($\pmb{\hat H}_{\text{SI}}\rightarrow\pmb{E}$) and $\pmb{\hat H}_\text{ext}$ tends to $\begin{bmatrix}\pmb{\hat H}_\text{DL}\\ \pmb{E}\end{bmatrix}$ so $\pmb{H}_\text{SI}$ and $\pmb{G}_\text{sps}$ tend to be independent. Thus, the expected SI power of the full duplex transceiver with the spatial suppression (the worst case) is given by
\begin{align}
{\bar \Omega}_{\text{sps}(\epsilon_{_\text{SI}}\rightarrow\infty)}
=\mathbb{E}\left\{||\pmb{w}_k^T\pmb{H}_\text{SI}\pmb{G}_\text{sps}||^2\right\}
=\mathbb{E}\left\{||\pmb{w}_k||^2\right\}.\label{sps_SI_w}
\end{align}
From (\ref{sps_SI_b}) and (\ref{sps_SI_w}), the approximation of  the expected SI power of the full duplex transceiver with the spatial suppression is expressed by
\begin{align}
{\bar \Omega}_\text{sps}\approx\frac{\mathbb{E}\left\{||\pmb{w}_k||^2\right\}}{1/\epsilon_{_\text{SI}}^2+1}.\nonumber
\end{align}
This approximation approach is a simple application of the harmonic mean.
Then, the approximation of the uplink SINR with the spatial suppression are given~by
\begin{align}
{\gamma}_{\text{sps}, k}^\text{UL}=\frac{K\rho_{_\text{UL}}^2(N-K)}{2K\rho_{_\text{UL}}+\frac{\rho_{_\text{SI}}}{\alpha_\text{anc}}\left(\frac{1}{1/\epsilon_{_\text{SI}}^2+1}\right)(K\rho_{_\text{UL}}+1)+1}.\label{SINR_sps}
\end{align}

\subsection{Comparison of SI-Subtraction and Spatial Suppression}
From (\ref{rate_stt}) and (\ref{rate_sps}), the SI-subtraction outperforms the spatial suppression in perfect channel estimation. Next, we compare the expected SI power of the spatial suppression and the SI-subtraction as follows: 
\begin{align}
{\bar \Omega}_\text{sps}
&=\mathbb{E}\left\{||\pmb{w}_k^T\pmb{E}\pmb{G}_\text{sps}||^2\right\}\nonumber\\
&=\mathbb{E}\left\{\sum_{n=1}^N\sum_{i=1}^K|w_{k,i}|^2\left(\sum_{m=1}^M|g_{k,m}|^2|E_{i,m}|^2+\zeta\right)\right\}\nonumber\\
&\le\left\{\sum_{n=1}^N\sum_{i=1}^K\mathbb{E}|w_{k,i}|^2\left(\sum_{m=1}^M\mathbb{E}|g_{k,m}|^2\mathbb{E}|E_{i,m}|^2+\bar{\zeta}\right)\right\}\nonumber\\
&=\epsilon_{_\text{SI}}^2\mathbb{E}\left\{||\pmb{w}_k||^2\right\}
={\bar \Omega}_\text{stt}\label{compSIC}
\end{align}
where the inequality is obtained by using Cauchy--Schwarz' inequality. $\zeta$ denotes $\sum_{m=1}^M\sum_{j=1}^M|g_{k,m}|^2|E_{i,m}^*E_{i,j}|$ and $\bar{\zeta}$~denotes $\sum_{m=1}^M\sum_{j=1}^M\mathbb{E}|g_{k,m}|^2\mathbb{E}|E_{i,m}^*E_{i,j}|$ where $w_{i,j}$, $g_{i,j}$, and $E_{i,j}$ are the $(i, j)$-th element of $\pmb{W}$, $\pmb{G}_\text{sps}$, and $\pmb{E}$, respectively.
From (\ref{compSIC}), we conclude that the ergodic performance of the spatial suppression is better than those of the SI-subtraction in imperfect channel estimation.

\begin{figure}[t]
 \centerline{\resizebox{0.9\columnwidth}{!}{\includegraphics{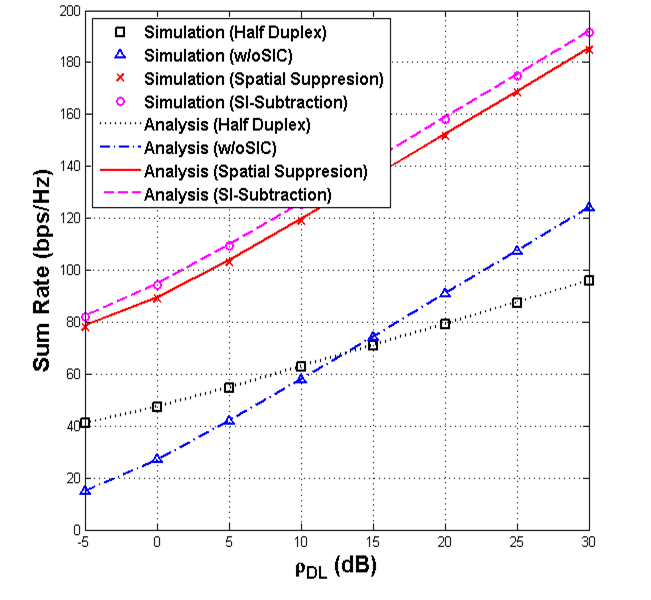}}}
  \caption{downlink/uplink sum rate vs. received downlink SNR with perfect channel estimation.}
  \label{Figure1}
\end{figure}

\section{Numerical Results}

For numerical comparisons, we assume that the full duplex-BS has a total of 84 antennas ($M=64$, $N=20$) and serves 10 users for each downlink and uplink. Simulation parameters are set as $\rho_{_\text{UL}}=10$~dB, $\beta_\text{SI}=-40$~dB, and $\beta_\text{UE} = -80$~dB, which are reproduced from the typical values of the path-loss at the reference meter (1~m) and those at small cell environments. We also assume that $\alpha_\text{acn}=40$~dB and $\text{NMSE}=0.2$. 

\subsection{Proposed Analysis and Simulation Results}

We first compare the proposed analysis and the simulation results. 
Figure~\ref{Figure1} shows that the results from (\ref{rate_noSIC}), (\ref{rate_stt}), and (\ref{rate_sps}) are approximately the same as the ergodic downlink/uplink sum rate of the full duplex-BS without the SIC and with the SI-subtraction/spatial suppression where perfect channel estimation is assumed at the BS. Since the SI is perfectly canceled out due to perfect knowledge of the SI channel, both uplink performance of the spatial suppression and the SI- subtraction are the same which use the same ZF receiver, i.e., there are a performance gap only in the downlink case. Additional spatial domain ($N$) is needed to cancel out the SI channel at the full duplex-BS with the spatial suppression, so we confirm that the SI-subtraction outperforms the spatial suppression in perfect channel estimation.

\begin{figure}[t]
 \centerline{\resizebox{0.9\columnwidth}{!}{\includegraphics{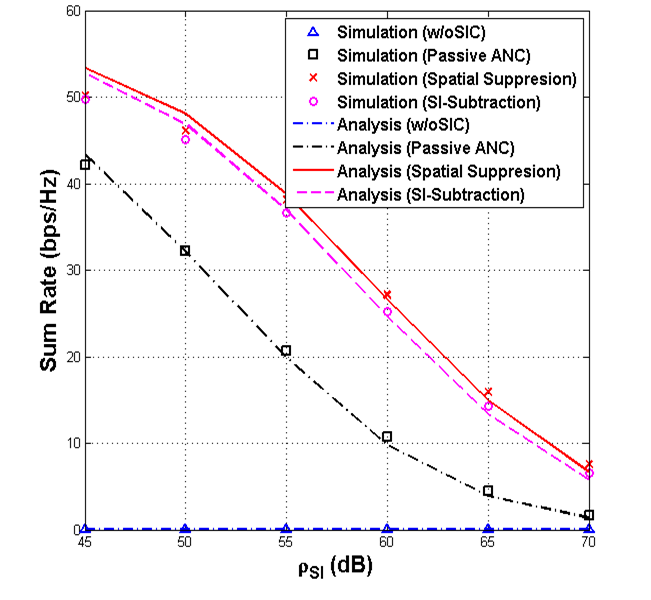}}}
  \caption{uplink sum rate vs. received SI SNR with imperfect channel estimation}
  \label{Figure2}
\end{figure}

Similarly, Fig.~\ref{Figure2} shows that the approximation forms, (\ref{SINR_noSIC}), (\ref{SINR_stt}), and (\ref{SINR_sps}), close to the ergodic uplink sum rate of the full duplex-BS without the SIC and with the SI-subtraction/spatial suppression where imperfect channel estimation is assumed. The legend, \emph{Passive ANC}, indicates the ergodic sum rate of the full duplex-BS only with a passive analog SIC and without any digital SIC. Note that these approximations are more tighter at the high received SI SNR regime, which are based on the approximation method for the ergodic sum rate in large-scale MIMO systems at the high SNR regime in \cite{Lim_TWC}. From the numerical results and the analysis, we can conclude that the spatial suppression technique could mitigate both SI term and estimation error term due to the correlation between the spatial suppression precoder and the estimation error of the SI channel.

\subsection{Simulation Results for a Correlated Channel}

Next, we compare the SI-subtraction and the spatial suppression in a highly correlated channel. To support huge data traffic demands, the full duplex BS is assumed to be located in a small cell, giving it compact antenna arrays with a high channel correlation.\footnote{A SDR platform for a small cell with compact multiple layers of a uniform linear array was presented in~\cite{Jang_WCM}.}
For numerical comparisons, the SI channel is a Rician channel modeled by the $N\times M$ matrix $\pmb{H}_\text{SI}$ with the line-of-sight (LOS) channel variance $\sigma_\text{SI}^2$ and Rician factor $\kappa$. Let $\pmb{R}_{\text{TX}}$ and $\pmb{R}_{\text{RX}}$ be spatial correlation matrices of the transmit antennas and the receive antennas at the BS.
From Jakes' model, the element of $\pmb{R}_{\text{TX}}$ and $\pmb{R}_{\text{RX}}$ is $r_{ij}^{\text{TX/RX}}=J_{0}\left({2\pi d_{ij}^\text{TX/RX}}/{\lambda}\right)$
where $d_{ij}^\text{TX/RX}$ is the distance between the $i$-th array and the $j$-th array of the transmit/receive antenna at the BS, and  $J_0$ and $\lambda$ are a zero-order Bessel function of the first kind and a carrier wavelength, respectively. Spatially-correlated DL, UL, and SI MIMO channels can be respectively modeled as: 
\begin{align}
\pmb{H}_\text{DL}&=\pmb{H}_{\text{DL}_\text{IID}}\pmb{R}_{\text{TX}}^{1/2},~~ \pmb{H}_\text{UL}=\pmb{R}_{\text{RX}}^{1/2}\pmb{H}_{\text{UL}_\text{IID}},\nonumber\\
\pmb{H}_\text{SI}&=\pmb{R}_{\text{RX}}^{1/2}\left(\sqrt{\frac{\kappa}{\kappa+1}}\sigma_\text{SI}+\sqrt{\frac{1}{\kappa+1}}\pmb{H}_{\text{SI}_\text{IID}}\right)\pmb{R}_{\text{TX}}^{1/2}\nonumber
\end{align}
where $\pmb{H}_{\text{DL}_\text{IID}}$, $\pmb{H}_{\text{UL}_\text{IID}}$ and $\pmb{H}_{\text{SI}_\text{IID}}$ are respectively i.i.d. Rayleigh DL, UL, and SI MIMO channels. The system operates at 2.1~GHz of the carrier frequency and imperfect channel estimation is assumed at the BS.

\begin{figure}[t]
 \centerline{\resizebox{0.9\columnwidth}{!}{\includegraphics{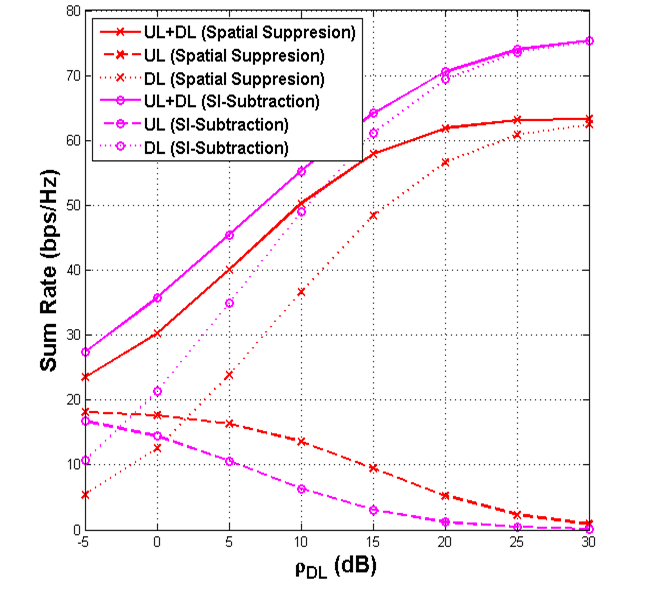}}}
  \caption{downlink/uplink sum rate vs. received downlink SNR with imperfect channel estimation where channels are correlated. More realistic parameters are used in simulation. The legends, \emph{UL} and \emph{DL}, indicate uplink and downlink, respectively.}
  \label{Figure4}
\end{figure}

 In Fig.~\ref{Figure4}, we consider a highly correlated SI channel with LOS components. Thus, a Rician channel is assumed with $\kappa=1$ and $\sigma_\text{SI}=1$, and each path-loss between the $i$-th receive array and the $j$-th transmit array is modeled by free-space path-loss according to $d_{ij}$ and the minimum value of $d_{ij}$ is $\frac{\lambda}{6}$ with a uniform linear array. The result shows that the uplink sum rate of the spatial suppression outperforms that of the SI-subtraction while the downlink sum rate of the SI-subtraction better than that of the spatial suppression. 

The more the downlink sum rates are increased, the more the uplink sum rates are degraded because high transmit power produce high SI power in a full duplex system. We should consider the performances at downlink and uplink, simultaneously. Figure~\ref{Figure4} shows that the overall sum rate of the spatial suppression outperforms that of the SI-subtraction for the given constraint of the uplink sum rate (e.g., 10 bps/Hz). Furthermore, the overall sum rate of the spatial suppression better than that of the SI-subtraction for the given constraint for the downlink sum rate where the uplink sum rates are guaranteed (e.g., at low $\rho_\text{DL}$ ($\le15$~dB)). We summarize these conclusions in Table~\ref{Table1}.

\section{Conclusion}
This paper has presented the performance analysis of the SI-subtraction and the spatial suppression in full duplex large-scale MIMO systems. The analysis and the numerical results have revealed the SI-subtraction outperforms the spatial suppression with perfect channel estimation because the spatial suppression requires additional antennas for null projection. We have also concluded that the ergodic performance of the spatial suppression is better than those of the SI-subtraction at uplink with imperfect channel estimation due to the correlation between the spatial suppression precoder and the estimation error of the SI channel. We provided an insight into which SIC algorithm is better under a given system constraint such as the downlink/uplink sum rate and the total transmit power at the BS. In future work, we will investigate the impact of the hardware impairments, user-to-user interference, and pilot overhead issues.

\section*{Acknowledgment}
This research was supported by the MSIP (Ministry of Science, ICT and Future Planning), Korea, under the ``IT Consilience Creative Program'' (IITP-2015-R0346-15-1008) supervised by the IITP (Institute for Information \& Communications Technology Promotion) and ICT R\&D program of MSIP/IITP.
\begin{table}[!t]
\caption{Desired SIC techniques for a given system constraint}
\begin{center}
\begin{tabular}{|c||c|}
\hline
Given System Constraint & Desired SIC Technique \\
\hline \hline
Downlink Sum Rate & Spatial Suppression \\
\hline
Uplink Sum Rate & Spatial Suppression \\
 \hline
Total Transmit Power & SI-Subtraction \\
 \hline
\end{tabular}
\end{center}
\label{Table1}
\end{table}
\bibliographystyle{IEEEtran}
\bibliography{FD_Lim_reference_v2}

\end{document}